# A Solution to the Temperature Evolution of Multi-well Free-energy Landscape

Yi Wang, Tiannan Yang, Shun-Li Shang, Long-Qing Chen, and Zi-Kui Liu

*Materials Science and Engineering, The Pennsylvania State University, University Park, PA 16802, USA.*

It has been a grand challenge to resolve the temperature evolution of multi-well free-energy landscape which is fundamentally relevant to phase transitions and associated critical phenomena as listed by Ginzburg [Rev. Mod. Phys. 76 (2004) 981]. To address this challenge, here we provide a simple solution based on *a priori* concept of Boltzmann thermal mixing among multiple parabolic potentials. The success and the impact of the present approach have been extensively demonstrated using a variety of materials, including Nb, $YBa_2Cu_3O_{7-\delta}$, $Ca_3Ru_2O_7$, Ni, $BiFeO_3$, and $PbTiO_3$.

# 1 Introduction

Phase transitions are the transition processes between different states of a medium, identified by the abrupt changes of some physical parameters – the critical phenomena. Systems with critical phenomena such as ferroelectric, magnetic, and superconductive phase transitions can mostly be treated as a problem of multi-well evolution across the free-energy landscape [1–7]. Meanwhile, it has been observed that many phase transitions exhibit similar behaviors [8,9] which are independent of the dynamical details of the system. This observation was termed as universality by Kadanoff [8]. As a result, most phase transitions can be theoretically formulated in the same manner, either using the Landau expansion [10,11] or using the piecewise polynomials [12].

The purpose of the present study is to develop an alternative theory to the widely adopted Landau expansion to resolve the problems of critical phenomena and phase transitions as listed by Ginzburg [11]. We show that the temperature evolution problem of a complicated free-energy landscape can be resolved by Boltzmann thermal mixing of multiple parabolic wells with different centers. The approach naturally gives rise to a long-range field and describes the free energy evolution from multi-well to single-well with increasing temperature. We will refer our approach as zentropy approach following the recent suggestion by Liu [13] on the approach of partition function method [14–16].

Let us define a reference system based on the concept of ground state. The ground state is at equilibrium and is characterized by a generalized vector $\Xi_0$ which can be such as the crystal orientation, the strain, the polar direction, the magnetization direction, the momentum, or the various combinations of them etc. Such a reference system is a pure state in the sense that the value of $\Xi_0$ is microscopically uniform everywhere. We then apply a virtual perturbation to the system to make it deviate from its equilibrium point $\Xi_0$ to an arbitrary point $\Xi$ (hereafter, $\Xi$ will be called the order parameter). Continually, we can write down a Taylor expansion to the second order for the Gibbs energy of the perturbed system.

$$\mathcal{H}(\Xi, T) = \mathcal{H}(\Xi_0, T) + \Delta\mathcal{H}(\Xi - \Xi_0) \qquad \text{Eq. 1}$$

where $\mathcal{H}(\Xi_0, T)$ is the Gibbs energy of the pure ground state at equilibrium, $T$ is the temperature, and

$$\Delta\mathcal{H}(\Xi - \Xi_0) = \frac{1}{2}(\Xi - \Xi_0) \cdot \left.\frac{\delta^2 \mathcal{H}}{\delta \Xi^2}\right|_{\Xi=\Xi_0} \cdot (\Xi - \Xi_0) \qquad \text{Eq. 2}$$

where $\frac{\delta^2 \mathcal{H}}{\delta \Xi^2}$ is a tensor representing the virtual second order derivatives of the Gibbs energy of the ground state with respect to the order parameter $\Xi$, and the dot symbol " $\cdot$ " represents the algebraic operation of dot product.

We propose that phase transition for a system is due to the microscopic disordering of the system in terms of $\Xi_0$. To do so, we postulate a macroscopic system made of a massive number of intrinsic elemental units (IEUs). Each of the IEUs can be distinguished by the orientations of their own $\Xi_0(\sigma)$, where $\sigma$ is the index of the individual IEU. We further assume that all $\mathcal{H}[\Xi_0(\sigma), T]'s$ introduced in Eq. 1 are independent of the orientations of $\Xi_0(\sigma)$, i.e., $\mathcal{H}[\Xi_0(\sigma), T]'s = \mathcal{H}_0(T)$ for all IEUs which is generally true when all the IEUs possess the same structure except for orientations. For the macroscopic system, the energy increase of each IEU can be expressed in the form of Eq. 2 using the macroscopic order parameter $\Xi$ and the microscopic order parameter $\Xi_0(\sigma)$. The partition function for the macroscopic system is then

$$Z = \sum_{\sigma} exp[-\beta \mathcal{H}^{\sigma}(\Xi, T)] \qquad \text{Eq. 3}$$

where $\beta = 1/k_B T$ is the Boltzmann constant and $T$ is the temperature, and the Gibbs energy can then be derived using $G = -\frac{lnZ}{\beta} = \mathcal{H}_0(T) + \Delta G(\Xi, T)$ where

$$\Delta G(\Xi, T) = -\frac{1}{\beta} ln \left( \sum_{\sigma} exp \left\{ -\beta \frac{1}{2} [\Xi - \Xi_0(\sigma)] \cdot \frac{\delta^2 \mathcal{H}}{\delta \Xi^2} \bigg|_{\Xi = \Xi_0(\sigma)} \cdot [\Xi - \Xi_0(\sigma)] \right\} \right) \qquad \text{Eq. 4}$$

$\Delta G(\Xi, T)$ accounts for the excess Gibbs energy. One realizes that $\Xi$ has the meanings of macroscopic field which is the thermal average of the individual $\Xi_0(\sigma)$.

Eq. 4 can be simplified when the second order derivatives of the Gibbs energy is a scalar, i.e., $\frac{\delta^2 \mathcal{H}}{\delta \Xi^2}\big|_{\Xi=\Xi_0(\sigma)} = \chi \mathbf{I}$ where $\chi$ is a scaler number and $\mathbf{I}$ is a unit matrix. As a matter of fact, when $T \to 0$, the excess Gibbs energy $\Delta G(\Xi, T)$ in Eq. 4 has multiple equivalent minima (2D) located at $\Xi_0(\sigma) = \pm X_0 \hat{x}_g$ ($g$ = 1, 2, …, D), where $\hat{x}_g$ is a unit vector. We further limit our consideration to the case of $\hat{x}_g \cdot \hat{x}_i = \delta_{gi}$ with $g$, $i$ = 1, 2, …, D. Here, while D is apparently the dimensionality of the system, it is closely related to the critical exponent of phase transition as will be demonstrated later in the present work.

As a proof of concept, we can consider the evolution of $\Delta G(\Xi, T)$ along a given representative direction, say along $\hat{x}_1$, then Eq. 4 can be simplified using $\hat{x}_1 \cdot \hat{x}_g = \delta_{1g}$

$$\Delta G(\xi, T) = Dk_B T_k \left\{ \frac{\xi^2}{2} + \frac{1}{2} - \frac{t}{D} ln \left[ 2(D-1) + 2cosh(\xi \frac{D}{t}) \right] \right\} \qquad \text{Eq. 5}$$

where $\xi = \Xi \cdot \hat{x}_1/X_0$, representing the reduced order parameter which would reach ±1 at 0 K and $T_k = \chi X_0^2/Dk_B$ is a characteristic temperature which is a reminiscent of the Curie temperature [17,18] in the Landau expansion [10,11]. It can be shown that $\frac{\partial^2 \Delta G(\xi,T)}{\partial \xi^2} = Dk_B \frac{T_k}{T}(T - T_k)$ at $\xi = 0$.

Figure 1 examines the general temperature dependences of the order parameter, the excess enthalpy, and the excess Gibbs energy. These thermodynamic properties have distinguished features between the first order and the second order transitions.

At $D = 1$, the present approach is able to recover the result of magnetization from the Weiss molecular field theory and the major results of free-energy from the Bragg-Williams approximation [19].

The case of $D = 3$ is a turning point, below which the phase transition is second order while above it the phase transition is first order. When $D \leq 3$, $T_k$ introduced in Eq. 5 equals the critical temperature of phase transition, i.e., for $T > T_k$, $\xi = 0$ corresponds to a global minimum; while for $T < T_k$ it is a local maximum. When $D > 3$, $T_k$ introduced in Eq. 5 becomes a precursor temperature which is slightly lower than the transition temperature.

As a generalization, $D$ can be treated with a value larger than 3 since it can be considered as an effective parameter to account for the other effects such as the coupled lattice distortions across the transition. At $D = 4$, $T_k$ introduced in Eq. 5 is no longer the critical temperature for the phase transition. Instead, the critical temperature increases from $t = T/T_k = 1$ to $t = 1.0567$ as given in Figure 1c for the case of excess free energy, showing triple wells at the critical temperature. For $D = 4$ and at the critical temperature, from Figure 1a one sees the sudden order parameter drop into zero, from Figure 1b one sees an abrupt jump in enthalpy – the latent enthalpy.

To exhibit different physics from different approaches, the excess Gibbs energies from the Landau expansion are also plotted in Figure 1d for comparison using PbTiO$_3$ as a prototype based on the parameters of Haun et al. [18]. The excess Gibbs energy modelled by the present approach shows a monotonic decrease with increasing temperature, i.e., a positive excess entropy which is a common sense of thermodynamics. In comparison, the Landau expansion shows a monotonic increase with increasing temperature due to its usage of the paraelectric state as the reference state, while the present approach uses the ground state as the reference state. The implication of this change of reference state is profound as the Gibbs energy of the ground state can be predicted from first-principles based phonon calculations, while the Gibbs energy of the paraelectric state cannot be accurately predicted at present due to the existence of imaginary phonon modes for most of the high temperature phases.

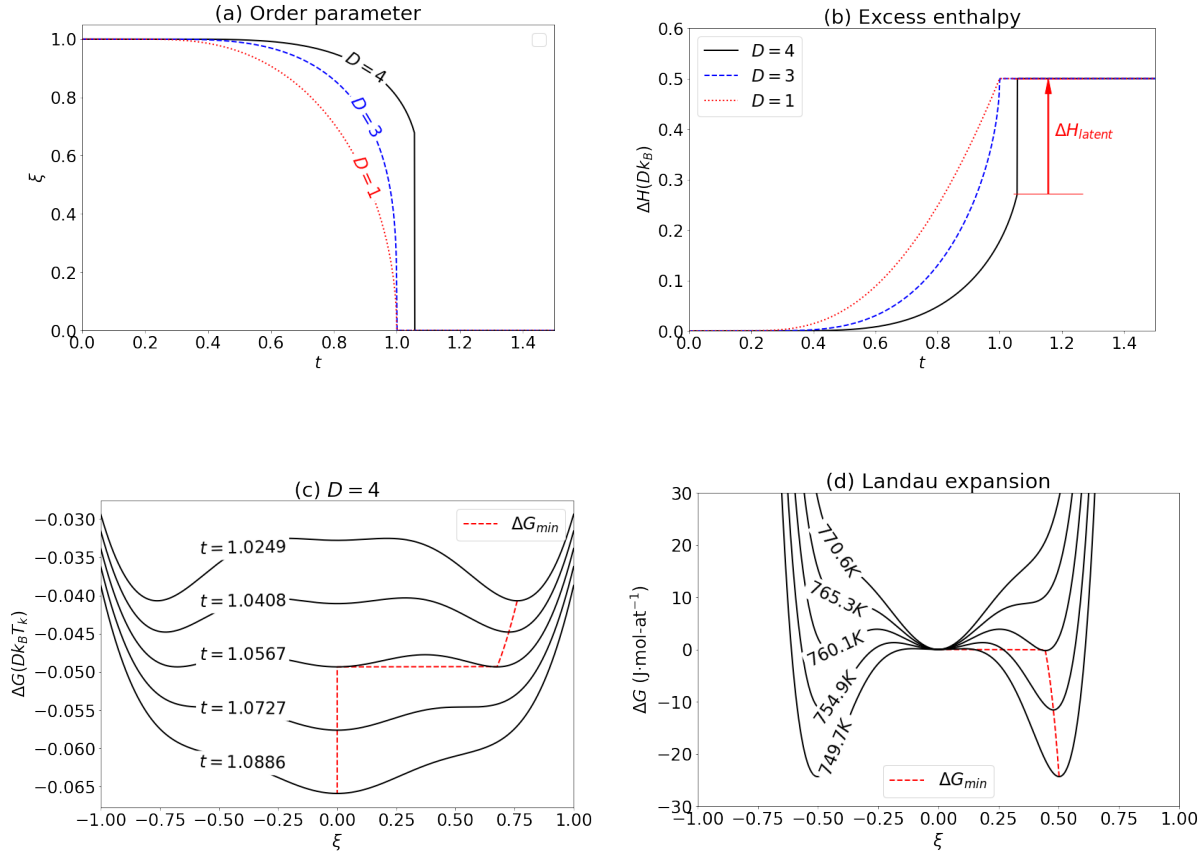

Figure 1. The temperature dependences of the excess thermodynamic properties. (a) the order parameter; (b) the enthalpy; (c) the excess Gibbs energy at $D = 4$; and (d) the excess Gibbs energy from the Landau expansion using parameters from Haun et al. [18] for PbTiO$_3$.

The order parameter $\xi$ plays the role of the effective *long range field* which effectively imposes the energy penalty when an IEU switches their directions. As a result, we can propose that the temperature dependence of $\xi$ is linearly correlated to, for example, the magnetization in magnetic transition, the pseudogap in superconductive or metal-insulator transition, and the polarization in ferroelectric transition. A salient feature of the order parameter commonly for almost all the phase transitions is that the order parameter exponentially vanishes when approaching to the critical temperature.

As a further generalization, to account for certain effects beyond the Cartesian dimension such as magnetism and superconductivity, we propose that $D$ can be treated as a non-integer fitting number so that it can be used to characterize critical exponent for the phenomena of interest. We find that the choice of $D = 2.5$ can universally describe the temperature evolutions of the corresponding characteristic order parameters for these 6 materials, seen by the comparisons with experiments [18,20–26] for the pseudo superconductive bandgaps for Nb and YBa$_2$Cu$_3$O$_{7-\delta}$ in Figure 2a and Figure 2b, respectively, metal-insulator transition for Ca$_3$Ru$_2$O$_7$ in Figure 2c, ferromagnetic-paramagnetic transition for Ni in Figure 2d, antiferromagnetic-paramagnetic transition for BiFeO$_3$ in Figure 2e, and ferroelectric transition in PbTiO$_3$. It is noted that as a conventional superconductor, even the BCS [27] theory has 15% deviation from experiment [21] for Nb as pointed out by French et al. [28], see Figure 2a. Noted also that for the spontaneous polarization vs temperature of PbTiO$_3$, the data points by Haun er al. [18] were from calculations based on the experimental spontaneous strain data, instead of the direct measurements.

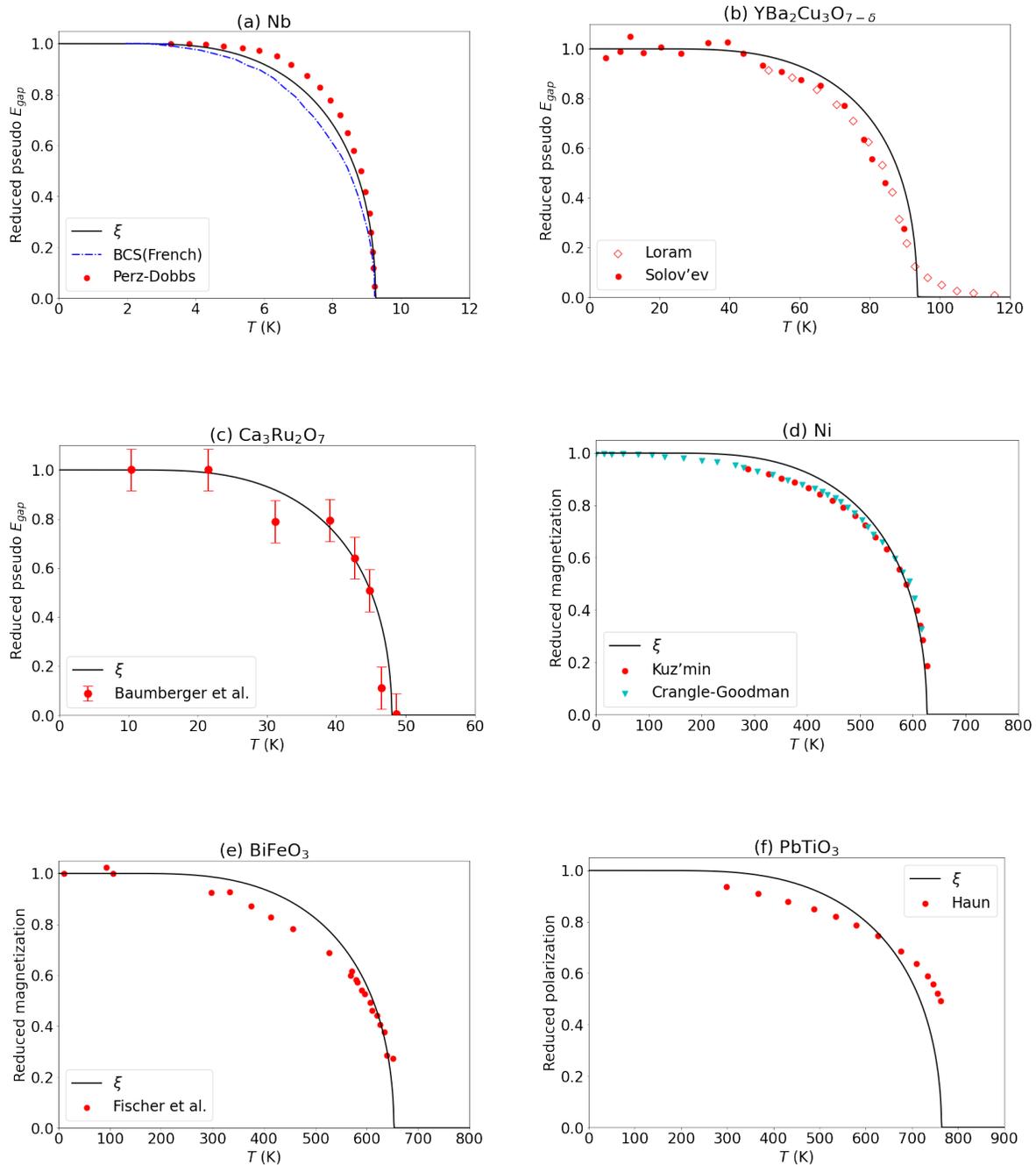

Figure 2. The calculated universal behaviors (lines) of the order parameters in comparison with experimental data (dots) for Nb [21] and BCS theory [28] [dot-dashed line in (a)), YBa$_2$Cu$_3$O$_{7-\delta}$ [29,30], Ca$_3$Ru$_2$O$_7$ (from the measurements by Baumberger et al. [23] at the momentum space point (0.5,0.5)), Ni [24,25], BiFeO$_3$ [26], and PbTiO$_3$ [18].

It is a common observation that phase transition can induce an excess heat capacity that mostly increases exponentially in wide range temperature range below the critical temperature. We find that the excess heat capacity can be well described by the present approach for a variety of phase transitions such as superconductive, magnetic, ferroelectric, and structural transitions as examined for the prototype materials of Nb, $YBa_2Cu_3O_{7-\delta}$, $Ca_3Ru_2O_7$, Ni, $BiFeO_3$, and $PbTiO_3$. To facilitate the comparison with experimental data, the phonon and electronic contributions as well as the effects of thermal expansion to the heat capacity are included using the results from our previous first-principles calculations for Ni [31] and $BiFeO_3$ [32], new first-principles calculations based on the projector-augmented wave (PAW) method implemented in the Vienna *ab initio* simulation package (VASP, version 5.2) using the DFTTK package [31] using the exchange-correlation (X-C) functional of GGA-PBE [33] for Nb, and the X-C functional of LDA [34] for $YBa_2Cu_3O_7$, and $PbTiO_3$, and $Ca_3Ru_2O_7$. To account for the effect of the dipole-dipole interaction on phonon frequency for insulators, the Born effective charge tensor and the high frequency static dielectric tensor needed in the mixed-space approach [35] are calculated by employing the linear-response theory as implemented in VASP 5.2 by Gajdos et al. [36].

Counting the contributions from lattice vibration, thermal electron, and the excess Gibbs energy, the total heat capacity per atom is

$$C(T) = \begin{cases} C_{vib}(T) + C_{el,n}(T) + \frac{\Delta C(T)}{N}, & for\ normal\ state \\ C_{vib}(T) + \frac{\Delta C(T)}{N} = C_{vib}(T) + C_{el,s}(T), & for\ superconductive\ state \end{cases} \qquad \text{Eq. 6}$$

where $C_{vib}(T)$ is the lattice contribution to the heat capacity, $C_{el,n}(T)$ is the thermal electronic contribution for normal metal state calculated based on the Mermin staticstics [37,38], and

$C_{el,s}(T) = \frac{\Delta C(T)}{N}$ is assumed by the present work for superconductive state with the excess heat capacity $\Delta C(T)$. Note that $N$ introduced in Eq. 6 is an effective number representing the number of atoms contained in the IEU, which is related to such as the critical size of superconductive domain (which may the size of the Cooper pair), magnetic domain, or ferroelectric domain etc.

The key parameters to calculate the excess heat capacity $\frac{\Delta C(T)}{N}$ are the dimensionality $D$, the characteristic temperature $T_k$ introduced in Eq. 5, and the effective size $N$ of the IEU introduced in Eq. 6. For the second order transitions, $T_k$ is set to the transition temperature. The principle to set $N$ is to use the integer multiplier of the number of atoms in the formula unit. Nevertheless, these quantities have been treated as fitting parameters in the present work. The used values for these parameters are collected in Table 1. It is found that $D = 2.5$ is a universal value to describe the superconductive and magnetic transition without structural change (no change of lattice shape) while $D = 4$ is a universal value to describe the transition coupled with structural change (including the lattice shape).

Table 1. Model parameters.

| Materials | Transition type | $D$ | $T_k$ | $N$ |
|---|---|---|---|---|
| Nb | Superconductive | 2.5 | 9.28 | 320 |
| YBa$_2$Cu$_3$O$_{7-\delta}$ | Superconductive | 2.5 | 93.6 | 91 |
| Ca$_3$Ru$_2$O$_7$ | Metal-insulator | 4 | 45 | 60 |

| Ca$_3$Ru$_2$O$_7$ | Antiferromagnetic-paramagnetic | 2.5 | 54.5 | 36 |
| Ni | Ferromagnetic-paramagnetic | 2.5 | 627 | 6 |
| BiFeO$_3$ | Antiferromagnetic-paramagnetic | 2.5 | 653 | 20 |
| BiFeO$_3$ | Ferroelectric-paraelectric | 4 | 1026 | 20 |
| PbTiO$_3$ | Ferroelectric-paraelectric | 4 | 725 | 10 |

Figure 3 shows the salient agreements between the calculations and experiments for the heat capacities in wide temperature range. The second order phase transitions, i.e., the superconductive phase transitions in Nb and YBa$_2$Cu$_3$O$_{7-\delta}$, the magnetic transitions in Ca$_3$Ru$_2$O$_7$, the magnetic transition in Ni, and the antimagnetic-paramagnetic transition in BiFeO$_3$, are shown as a heat capacity spike or discontinuity (but finite values) across the transition temperatures. The first order phase transitions, i.e., the metal-insulator transition in Ca$_3$Ru$_2$O$_7$, the ferroelectric transition in BiFeO$_3$, and the ferroelectric transition in PbTiO$_3$, are shown as an infinite heat capacity jump across the transition temperatures. It is very interesting to note that the modeled heat capacity spikes by the present approach agrees very well with the measurements for the superconductive transitions [29,39–41], for example, the steep heat capacity drop across the superconductive phase transitions for Nb in Figure 3a and for YBa$_2$Cu$_3$O$_{7-\delta}$ in Figure 3b. In particular, the calculated electronic specific heat coefficients plotted in the inset of Figure 3b agrees very well with the measured data for YBa$_2$Cu$_3$O$_{6.92}$ by Loram et al. [29], noticing the fact that there is still a need for the theory on the high temperature superconductors (represented by the cuprates).

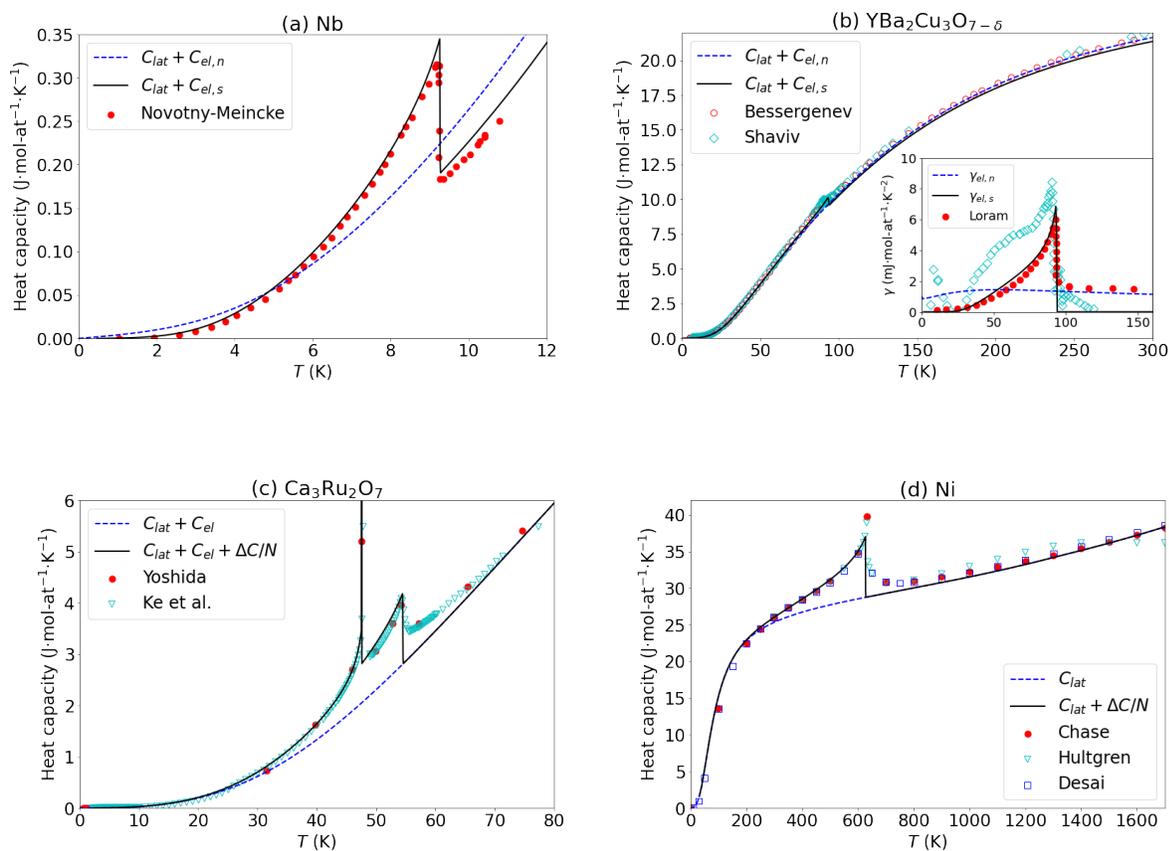

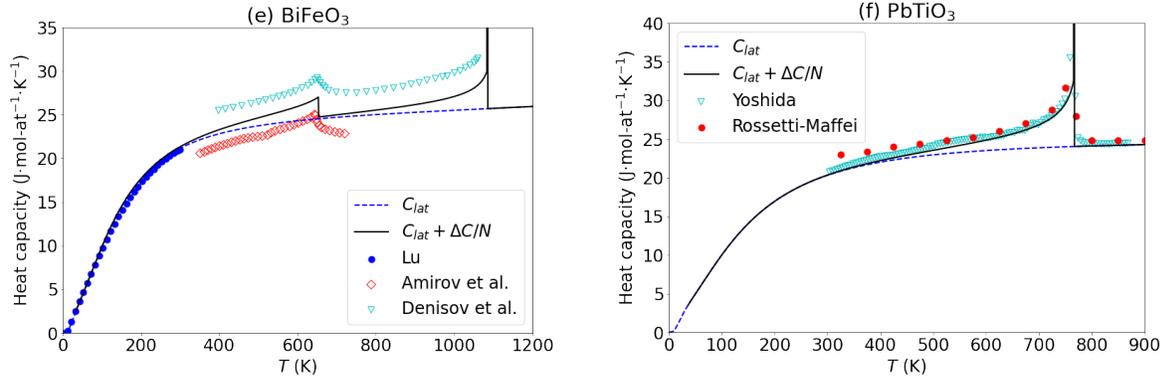

Figure 3. Calculated heat capacities (lines) in comparison with experimental data (dots) for Nb [39], YBa$_2$Cu$_3$O$_{7-\delta}$ [29,40,41], Ca$_3$Ru$_2$O$_7$ [42,43], Ni [44–46], BiFeO$_3$ [47–49], and PbTiO$_3$ [17,50]. The inset of in (b) shows the electronic specific heat coefficients for YBa$_2$Cu$_3$O$_{7-\delta}$.

In summary, the present work demonstrates an approach for handling the general critical phenomena and phase transitions by Boltzmann thermal mixing among multiple parabolic potentials. We show that a macroscopic system with a complex free-energy landscape can be decomposed into an ensemble of multiple local parabolic wells. The approach is able to account for the lattice vibration and is applied to a variety of phase transitions, including superconductive, magnetic, ferroelectric, and structural transitions as exemplified using Nb, YBa$_2$Cu$_3$O$_{7-\delta}$, Ca$_3$Ru$_2$O$_7$, Ni, BiFeO$_3$, and PbTiO$_3$. The key findings include: i) the excess free-energy is formulated using only the second order term of the order parameter, in compared four or higher orders used by the Landau expansion; ii) the long range field is derived as *a posteriori* result; iii) the MPP approach naturally brings out the double-well and its evolution to single-well, being in comparison with the Landau expansion that is basically *a priori* control for the coefficient of the second order term, making it positive above the phase transition temperature and negative below the phase transition temperature; iv) the excess free-energy is referenced to the ground state structure whose free-energy can be well formulated and calculated within the framework of the first-principles theory, compared with that of Landau expansion referenced to high temperature structure for which the issues of imaginary phonon modes cannot be avoided for most materials; v) for magnetic system, the Weiss formula for magnetization can be derived straightforward; vi) a single choice of the critical exponent (dimensionality) $D = 2.5$ can universally describe the temperature evolution of the corresponding characteristic order parameters as demonstrated for the aforementioned six materials; vii) For heat capacity, $D = 2.5$ is an universal value for describing the superconductive and magnetic transitions, while $D = 4$ is an universal value for describing the transition associated with structural change.

## Acknowledgments

This work was mainly supported from the Computational Materials Sciences Program funded by the US Department of Energy, Office of Science, Basic Energy Sciences, under Award Number DE-SC0020145 (Wang, Yang, and Chen); and partially supported from the National Science Foundation (NSF) with Grant No. CMMI-2050069 (Wang, Shang and Liu). First-principles calculations were performed partially on the Roar supercomputer at the Pennsylvania State University's Institute for Computational and Data Sciences (ICDS), partially on the resources of the National Energy Research Scientific Computing Center (NERSC) supported by the U.S. DOE Office of Science User Facility operated under Contract No. DE-AC02-05CH11231, and partially on the resources of the Extreme Science and Engineering Discovery Environment (XSEDE) supported by NSF with Grant No. ACI-1548562.